\documentclass[conference]{IEEEtran}
\IEEEoverridecommandlockouts
\usepackage{cite}
\usepackage{amsmath,amssymb,amsfonts}
\usepackage{algorithmic}
\usepackage{graphicx}
\usepackage{textcomp}
\usepackage{xcolor}
\usepackage[ruled,vlined]{algorithm2e}
\usepackage{url}
\usepackage{comment}
\usepackage{balance}
\usepackage{caption}
\usepackage{subfigure}
\usepackage{cleveref}
\usepackage{tikz}
\usepackage{textcomp}
\usepackage{lipsum}

\begin{document}

\newenvironment{conditions}
  {\par\vspace{\abovedisplayskip}\noindent\begin{tabular}{>{$}l<{$} @{${}={}$} l}}
  {\end{tabular}\par\vspace{\belowdisplayskip}}

\newcommand\copyrighttext{%
  \footnotesize \textcopyright 2020 IEEE. Personal use of this material is permitted.
  Permission from IEEE must be obtained for all other uses, in any current or future
  media, including reprinting/republishing this material for advertising or promotional
  purposes, creating new collective works, for resale or redistribution to servers or
  lists, or reuse of any copyrighted component of this work in other works.
  }
\newcommand\copyrightnotice{%
\begin{tikzpicture}[remember picture,overlay]
\node[anchor=south,yshift=10pt] at (current page.south) {\fbox{\parbox{\dimexpr\textwidth-\fboxsep-\fboxrule\relax}{\copyrighttext}}};
\end{tikzpicture}%
}

\title{PiChu: Accelerating Block Broadcasting in Blockchain Networks with Pipelining and Chunking}

\author{{Kaushik Ayinala \hspace{0.3cm} Baek-Young Choi \hspace{0.3cm} Sejun Song\hspace{0.3cm}}\\
{University of Missouri-Kansas City, Kansas City, MO, USA }\\
Email: \{kapnb, choiby, songsej\}@umkc.edu}

\maketitle

\copyrightnotice

\begin{abstract}
Blockchain technologies have been rapidly enhanced in recent years. 
However, its scalability still has limitations in terms of throughput and broadcast delay as the network and the amount of transaction data increase.  
To improve  scalability of blockchain networks, we propose a novel approach named PiChu that accelerates block propagation in blockchain networks by pipelining and verifying chunks of a block in parallel. 
Accelerating block propagation reduces the mining interval and chance of fork occurring, which in turn increases throughput. Our approach can be applied to the blockchain networks either directly or with a minor modification to the consensus. 
Through an extensive and large scale simulations, we validate that the proposed PiChu scheme significantly enhances the scalability of blockchain networks. For instance, a 64 MB block can be broadcasted in just 80 seconds in a blockchain network with a million nodes. The efficiency of PiChu broadcasting increases with bigger block sizes and a larger number of nodes in the network.
\end{abstract}

\begin{IEEEkeywords}
blockchain, block propagation, chunking, pipelining, simulator, P2P network, scalability
\end{IEEEkeywords}

\section{Introduction} \label{sec:intro}

Blockchain maintains a distributed ledger of the completed transactions as blocks and chains them sequentially using the previous block hash to maintain the order of completed transactions.  Nodes in a blockchain are connected to each other on a peer-to-peer (P2P) network. A consensus protocol running at every node follows the agreement of a policy to add a block to the chain.

There are several consensus schemes such as Proof of Work (PoW), Proof of Stake (PoS), Delegated Proof of Stake (DPoS), Practical Byzantine Fault Tolerance (PBFT), and Hybrid Consensus. 
For instance, Proof of Work (PoW) is one of the commonly used consensus algorithms introduced in bitcoin~\cite{bitcoin}.
In the PoW consensus algorithm, each block contains a timestamp, nonce, hash of the block, difficulty target. Proof of Stake (PoS) is another well-known consensus algorithm introduced in PPCoin \cite{king2012ppcoin}.

After validating a block, the node broadcasts or propagates it to the rest of the network. The time it takes to propagate a block depends on many factors, such as the size of a block, the average bandwidth of the nodes, and the maximum hop count or diameter of a network. Those factors have intricate relationships.
When the number of nodes in the network increases,  the network diameter increases along with a block broadcast time.  Also, when throughput is increased via larger block size, a block broadcast time increases, causing the chance of undesirable forks. The blockchain network becomes unstable when there are too many forks, or forks do not resolve. Therefore, if we increase the throughput or capacity of the blockchain network, then it may become unstable. This causes the scalability problem~\cite{SecureSharding, BlockPerformance, BitcoinP2PNetwork} in the blockchain.

In this paper, we propose a Pipelining and Chunking scheme for blockchain networks, named PiChu that is to expedite a block propagation by verifying consensus with the block header and incrementally forwarding the body of a block by small chunks  over the P2P network, instead of a whole block at once. After receiving a chunk, a node will verify and forward the chunk. Accelerating block broadcast time improves the scalability of the blockchain, as the block interval can be reduced, the block size can be increased, and forks in the chain would be reduced.

Since PiChu takes advantage of network pipelining, the efficiency is far better than the traditional approach. Our experimental results showed, on average, a magnitude ($\approx$ 13.6 times) less block broadcast time than the traditional method in a blockchain network with 65,536 nodes. PiChu technique can be applied directly to the existing consensus protocols with  minimal change to the blockchain network. 
PiChu approach can be directly used with the existing consensus algorithms such as PoS, DPoS and PBFT that use a header only to verify a block. 
As for PoW that uses an entire block for a verification, PiChu approach can be employed with a minor modification in the consensus.  

Our contributions in this paper include
i) proposing PiChu, a novel block broadcasting technique,
ii) development of a versatile blockchain simulator, and  
iii) analysis and extensive evaluations of the efficiency of the proposed scheme.

The rest of the paper is organized as follows. We discuss the existing works on blockchain scalability in Section~\ref{sec:relatedwork}.  Section~\ref{sec:scheme} describes the proposed scheme in detail. The efficiency and the pseudo-code of our scheme are given in Section~\ref{sec:efficiency}. Section~\ref{sec:security} discusses the potential attacks and proposes countermeasures. Section~\ref{sec:experiments} explains the experiment environment and results. We conclude the paper in Section~\ref{sec:conclusion}.

\section{Related Work} \label{sec:relatedwork}
There are a number of studies to improve the scalability of the blockchain networks. They follow approaches like using multiple chains, sharding or exploiting  network topology information.

Monoxide~\cite{monoxide} uses multiple chains to linearly scales the blockchain. It proposes Chu-ko-nu mining to maintain the same difficulty across all the chains, and proposes a protocol to handle inter-chain transactions. A node can mine a block in multiple or all chains by solving a single problem. Miners can choose the chains they want to work on. This may cause a chain to be abandoned if there are too many chains. Elastico~\cite{SecureSharding} also linearly scales the blockchain by  sharding but uses a single chain. Sharding involves dividing the network into groups or shards for a given amount of time. Each group will work on a different set of transactions. The size of the block increases with the number of nodes, which in turn increases the broadcast time. The size of the block is limited by bandwidth and latency.

A scheme to speed up block propagation by choosing the closest neighbors as peers was proposed in~\cite{ACMMP}, where the closest neighbor is determined by  transmission latency. Another study~\cite{BBBP} also improves the scalability by maintaining   the network topology using a  tree structure for a broadcast routing.  Tree cluster routing is proposed to do routing during node failures. However, it does not address adapting to dynamic network conditions such as  a new node's join and  handling a node or cluster failure. Velocity~\cite{velocity}  improves block broadcasting by downloading the parts of a block from multiple neighbors.
In the scheme, a block is  converted into so-called fountain codes. The node that wants to receive a block sends a request message to all of its neighbors. The neighbors having the block sends a fountain code continuously. After receiving sufficient codes, the node rebuilds the block.  Graphene~\cite{Graphene} improves block propagation by reducing the transmission delay between the nodes. Graphene uses Bloom filters and Invertible Bloom Lookup Table (IBLT) to synchronize the block between peers.
Bitcoin-NG~\cite{Bitcoin-NG} indirectly selects a leader for a given time frame, and the leader transmits the micro blocks throughout the time frame. The chain contains two types of blocks. They are key and micro blocks. The node that mines the key block becomes the leader. The consensus protocol for the key block is PoW. However, it is for a specific type of a consensus protocol and can not be used on other existing consensus protocols.  

To the best of our knowledge, this paper is the first work that uses the unique approach of pipelining and chunking for accelerating block propagation blockchain networks.  The proposed scheme can be used along with existing scaling and acceleration techniques in a complementary manner.

\section{PiChu: The Proposed Pipelining and Chunking Approach} \label{sec:scheme}

This section explains the proposed PiChu scheme. PiChu scheme involves first, sending a header as an invitation, then dividing the body of the block into chunks, and finally forwarding the chunks in pipeline.

\subsection{Verification of a Block for a Consensus}

PiChu does a block verification for a consensus using only a header rather than a whole block.  

Most consensus algorithms including Proof Of Stake, Delegated Proof of Stake, Proof of Activity, Proof of Burn, Proof of Elapsed Time, and Leased Proof of Stake (\cite{king2012ppcoin, Novacoin, consensuses, Consensusesurl, ferdous2020blockchain}) need only the header for the consensus verification. Those consensus protocols require only the header to verify  a block for the consensus, as shown in Equation~(1). Equation~(1) is the consensus between the nodes in PoS. Thus PiChu can be readily used on those blockchains.  

\begin{equation}
\label{equation:PoS}
    Hash(Header) < C_w* \textit{DifficulyTarget}
\end{equation}
where,
$C_w$ is a coin day weight.

On the other hand, Proof of Work is a consensus protocol that requires an entire block for its verification. However, it can be made PiChu-capable with  minor modifications.  Note that nodes in a PoW blockchain follows  Equation~(\ref{equation:PoW}) to add a block to their chain.
\begin{equation}
   Hash(Block) < \textit{DifficulyTarget}
   \label{equation:PoW}
\end{equation}

Hash of all the transactions should be included in the header. The reward transaction should be included in the header. In Bitcoin, the size of nonce is 32 bit, but the difficulty target can be greater than $2^{32}$. Miners iterate nonce, but they may not find the nonce that satisfies the consensus, and then miners shuffle the transactions and iterate the nonce again. As consensus has to be verified with the header, there are no transactions for the miners to shuffle. So the size of the nonce has to be increased. After modifying the header, the PoW consensus can be verified by using the Equation~(\ref{equation:CTPoW}).
The PoW consensus is modified to use only the header. PiChu can now be used on modified PoW consensus.
 \begin{equation}
  \label{equation:CTPoW}
    Hash(Header') < \textit{DifficulyTarget}
\end{equation}

\subsection{Chunking and Propagation Scheme}
Chunking involves dividing the body of the block into multiple chunks of the same size. Each chunk should contain only complete transactions, and the remaining space in a chunk is padded.

A miner appends some information about chunking to the header and signs it. The miner can not use any key to sign the header with metadata. He has to sign with the key that is used to claim the reward of the block. All the blockchains give rewards to the miners. The reward is included in the block. The reward should contain the public key of the miner. In consensus algorithms that need a whole block, the reward is included in the body of the block.  We have to modify a whole-block consensus protocol in such a way that the reward for mining the block should be included in the header. Thus, a miner signs the header with metadata by using its reward private key. Miner sends the signature along with the header to its connection.  Receiving nodes retrieves the header from the invitation and verifies the consensus. If consensus is correct, then nodes retrieve the miner reward public key and verify the signature of the invitation. If it is correct, then the node retrieves the information about the chunks from the invitation and uses it receives the chunks. After dividing the block into chunks, the miner appends the chunk number at the starting of the chunk to identify the order of the chunks. The miner then signs the chunk with metadata by using the reward private key, and also signs each chunk to prevent an intermediate node from tampering the data. When a node receives a chunk, it checks the integrity by using the reward public key. We discuss about the optimal chunk size in  Section~\ref{sec:efficiency}.

\begin{figure}
\centering 
  \includegraphics[width=0.95\linewidth]{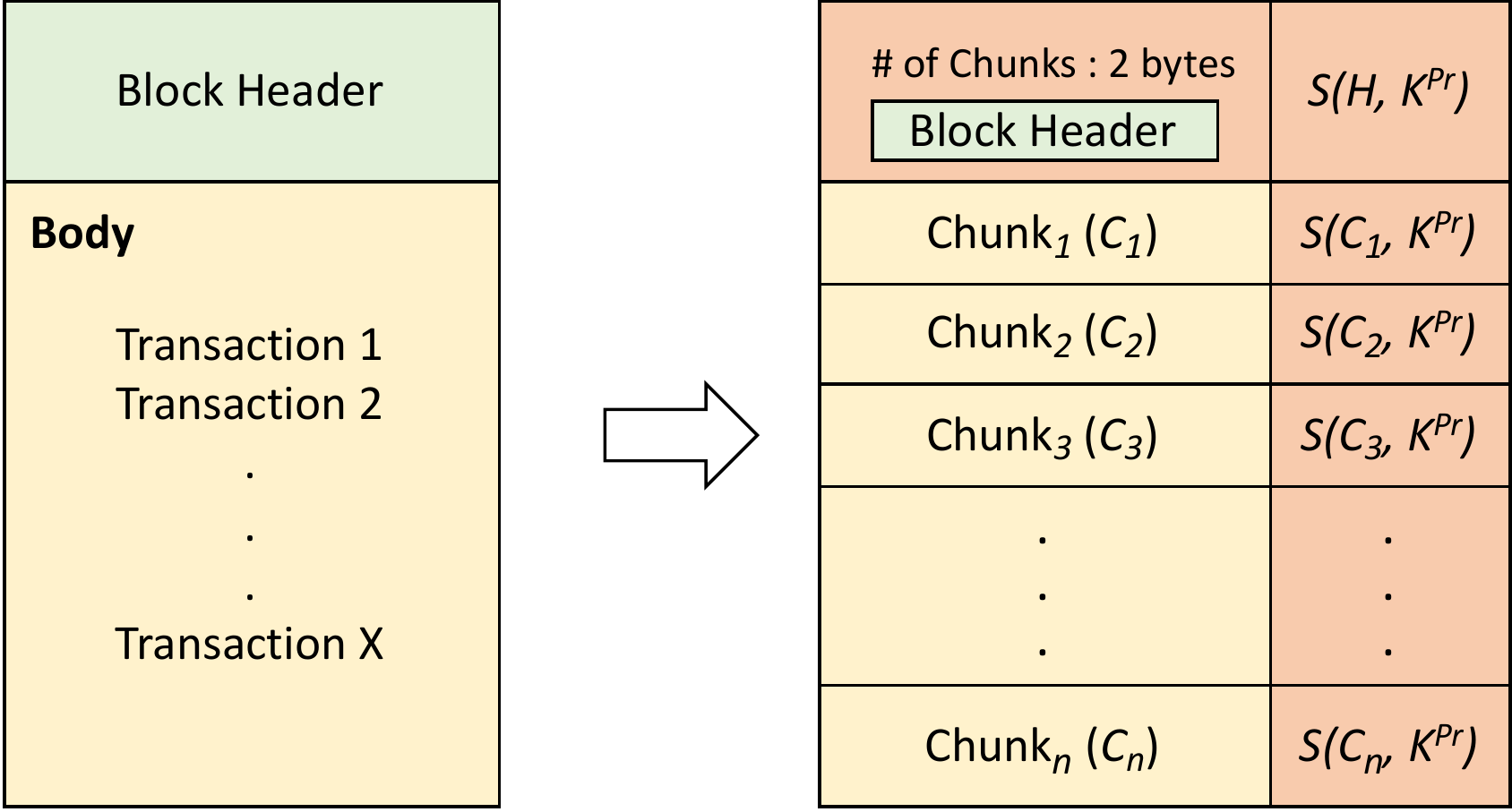} 
  \caption{Block structure in PiChu}
  \label{fig:pichublock}
\end{figure}
\begin{table}
 	\centering
 	\caption{PiChu field types description}
 	\label{tab:table1}
 \begin{tabular}{|c|c|p{4cm}|}
 
 		\hline
        Field Name & Size & Description\\
 		\hline \hline
 		\# of chunks & 2 Bytes & Number of chunks in the body of block, varies with number of transactions in the body \\\hline

 		$C_i$ & 128 KBytes &$i^{th}$ chunk in the body of the block\\ \hline
 		
 		$S(C_i, K^{Pr})$ & 64 Bytes & signature of  $C_i$  with ECDSA private key in the block header \\ \hline
 		
 	\end{tabular}
 \end{table}

For every chunk, we send an additional 64 bytes as a signature, which increases the amount of data to be transferred for a block. Even though the data transmission size is increasing, the blocks are transmitting much faster. While storing the block, nodes can remove the metadata about the chunks.

\begin{figure*}[t]
\centering
\begin{minipage}{0.48\textwidth}
\centering
\includegraphics[width=0.99\linewidth, height=0.25\textheight]{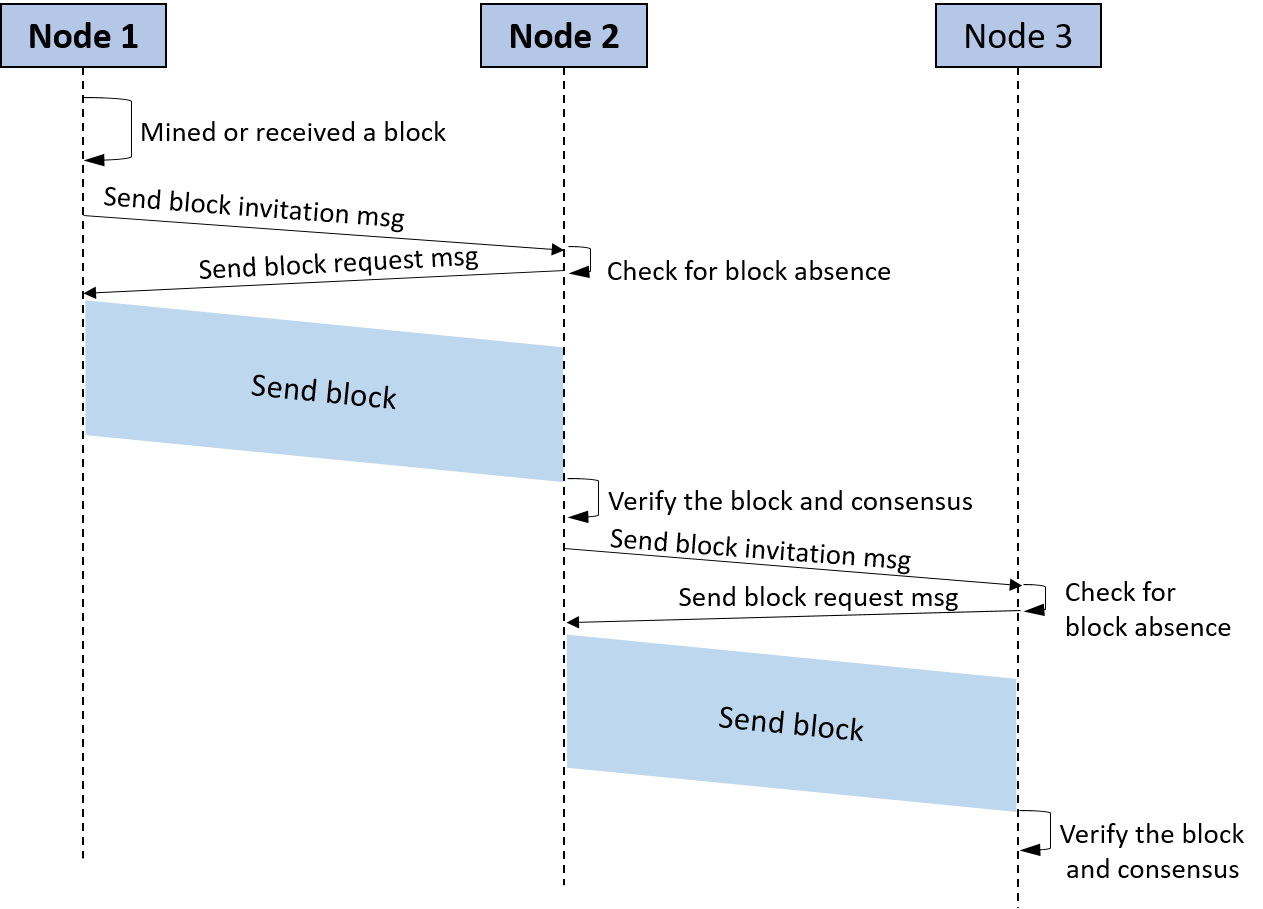}
\caption{Block broadcast sequence in traditional blockchains}
  \label{fig:generalsequence}
\end{minipage}
\begin{minipage}{0.48\textwidth}
\centering
\includegraphics[width=0.99\linewidth, height=0.25\textheight]{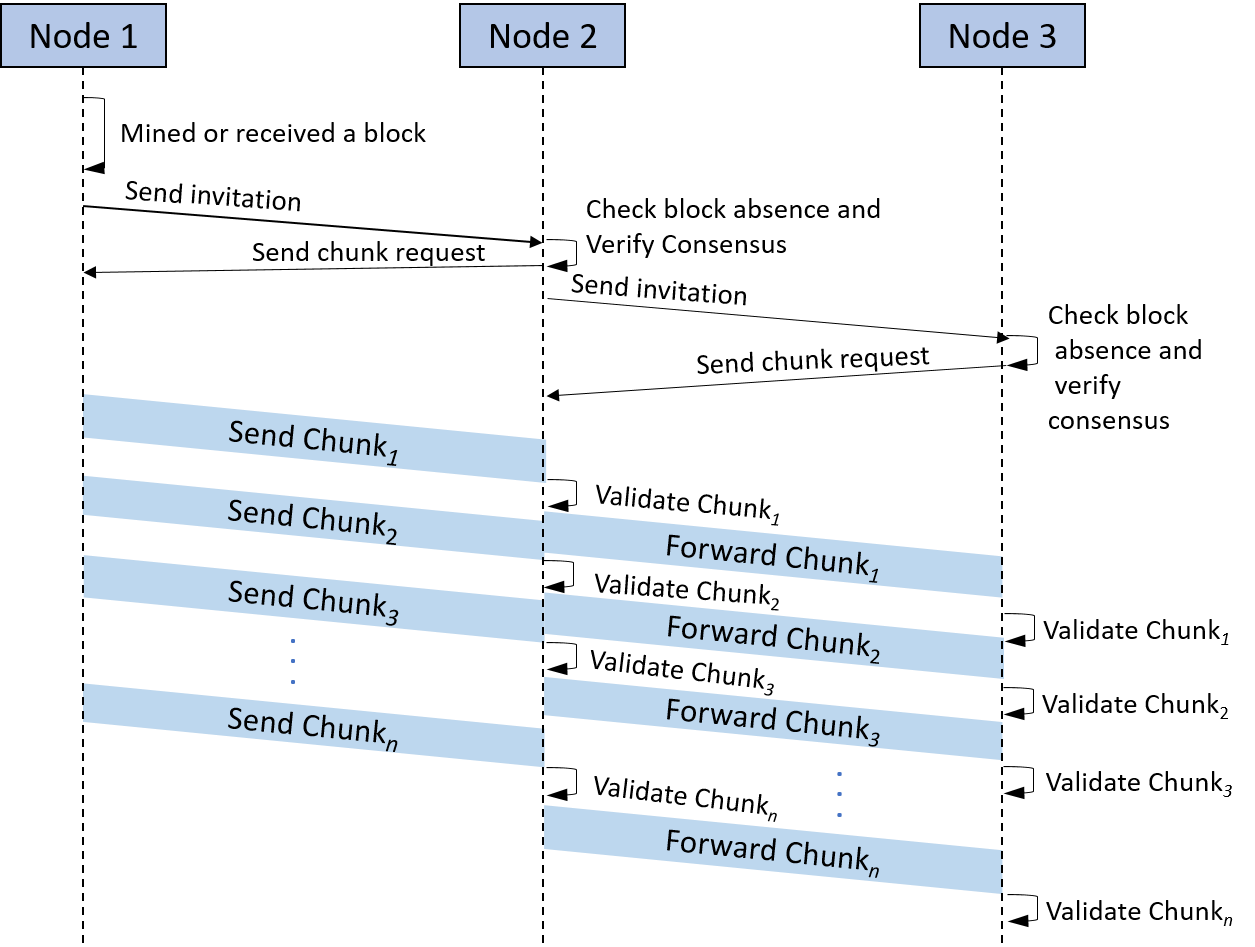}
\caption{Block broadcast sequence in PiChu blockchain}
  \label{fig:pichusequence}
\end{minipage}
\end{figure*}
\subsection{Pipelining}

In the general broadcast approach, when a node mines or receives a new block, it sends a block invitation to all of its neighbors. If a node receives a block invitation, it checks whether block exists or not. If the block does not exist, then the node replies with the block request message. After receiving the block request message, the node forwards the block. The receiving node verifies the block, and if the block is valid, then it sends the block invitation to all of its neighbors. The traditional block broadcast protocol is illustrated in  Figure~\ref{fig:generalsequence}.

As illustrated in Figure~\ref{fig:pichusequence}, when a node mines or receives a block, it sends an invitation to all the connected nodes with the PiChu header. The node that received the invitation message verifies whether the header achieves consensus or not. If the header achieves consensus and the node does not have that block, it sends a chunk request message back to the original node.

Besides, it sends an invitation message to its neighbor nodes by using the PiChu header. After sending the header invitations to all the neighbors, the miner starts sending the chunks to the neighbors who sent the chunk request message. When a node receives chunks, it verifies the signature of each chunk by using the public key of the miner. Although an additional 64 bytes as a signature is required for each chunk, the overhead is trivial. As long as a chunk is verified, it forwards the chunk to its neighbor nodes, which sent a chunk request. The verification of chunk includes checking the integrity and validity of transactions in it.

\section{Analysis of PiChu Efficiency}\label{sec:efficiency}
The broadcast time is proportional to the radius of the network. If the network radius increases, then broadcast time increases. The broadcast time is also proportional to the delay at each node. If the delay at each node increases, then block broadcast time increases and vice versa. 

On average, the broadcast time in a traditional blockchain network is equal to the radius of the network in hop counts multiplied by the nodal delay of a network, and the nodal delay is the sum of the transmission delay, propagation delay, and verification delay. The transmission delay depends on the block size, bandwidth, and the number of neighbors. The transmission delay is proportional to block size and the average degree of the nodes. The transmission delay is inversely proportional to the bandwidth. The verification delay also depends on the size of the block and the diameter of the blockchain network. The notations of the symbols used in this section are summarized in Table~\ref{tbl:NotationExplanation}.

\begin{eqnarray}
T_B   &=& R \times \left(T_{LinkTrans} + T_{LnkPrp} + T_{ver} \right)\\
   &= & R \times\left(D_{conn} \times \frac{L_B}{B_w}  + T_{LnkPrp} + T_{ver}\right) \label{equation:generalprop}
\end{eqnarray}

In the PiChu scheme, the header is broadcasted first, then chunks are pipelined in parallel. So the time it takes to propagate the block is equal to the sum of the time it takes to broadcast the header and the time to transmit all the chunks from a node to another. The time it takes to transmits chunks from one node to another depends on the degree of the nodes, the number of chunks, metadata, and bandwidth. The size of metadata for each is 520 bits.

 \begin{eqnarray} 
   T_{PiChu} &= &T_{PH} + T_{DC}\\
 &  = &R \times \left ( D_{conn} * \frac{L_H}{B_w}  + T_{LnkPrp} + T_{ver}\right) + T_{DC}\nonumber \\ 
 & = & R\times\left( D_{conn} \times \frac{L_H}{B_w}  + T_{LnkPrp} + T_{ver}\right) + \nonumber  \\ 
  & & \frac{D_{conn} \times \left(N_c+520\right)\times L_C}{B_w}  \label{equation:pichuprop}
 \end{eqnarray}

As seen in Equation~(\ref{equation:generalprop}), the block broadcast time depends on the product of the network radius and block size. If the block size is increased in Equation~(\ref{equation:generalprop}), then broadcast time increases by at least $R$ times. In Equation~(\ref{equation:pichuprop}), we can observe that the block broadcast time depends on the product of the network radius and header size. If the block size is increased in Equation~(\ref{equation:pichuprop}) then broadcast time increases by only block transmission delay between two nodes. So the PiChu block broadcast approach is very efficient than the general broadcast approach. The efficiency of the PiChu broadcast approach over the traditional broadcast approach increases with an increase in block size and number of nodes in the network. 

Algorithm \ref{algo:scheme} gives the pseudo-code of the PiChu scheme. It shows that when a node receives an invitation from a peer, it checks whether that header or block exists in the chain. If it does not exist, then the node requests and receives the chunks from the peer. When a chunk is received, it immediately forwards it to other peers. It also indicates that only one block is received and forwarded at a time.\\

\begin{algorithm}
\SetAlgoLined
List HeaderConnections; Object CurrentHeader\;
 \While{True}{
  Receive block header H as an invitation from node N\;
  \uIf{CurrentHeader == H}
  {HeaderConnections.add(N)\;
    Continue\;
  }
  \uElseIf {CurrentHeader == null}
  {
    CurrentHeader = H\;
    \uIf{If adding H makes a chain longest}{
        \uIf{Hash(Header)\textless  D}{
        sendToOthers(H)\;
        Retrieve $Pu_k$, $N_c$ from H\;
        
        \While{$N_C-- > 0$}{
            Receive a chunk\;
            
            \uIf{Chunk is valid}{
                sendToOthers(Chunk)\;
            }
            \uElse{
                Choose a node X from HeaderConnections\;
                $N_C ++$ \;
                Request X to pipeline last $N_C$ chunks\;
            }
        }
        \uIf{Block is valid}{
            Add block to the chain;
        }
        \uElse{
            Discard the block;
        }
    }
    }

  }

 }
 \SetKwFunction{sendToOthers}{sendToOthers}
 \SetKwProg{myproc}{Procedure}{}{}
  \myproc{\sendToOthers{Data}}{
  
    Send Data to other nodes in parallel
    
  }

 \caption{Psuedo code of Chunking and Pipelining Block Broadcast Scheme }
 \label{algo:scheme}
\end{algorithm}

The size of the chunk is bounded by the block size but should be large enough to overcome the metadata processing overhead. The transmission delay of a chunk should be less than the sum of propagation delay and protocol overhead so that there will be no extra delay at each forwarding node.  A node has to receive the chunk before it receives the chunk request message from its neighbors so that it can immediately forward the chunk after receiving the message.
The chunk size can be decided by Equation~(\ref{equation:chunksize}) below.

\begin{eqnarray}
    T_{tc} &<& T_{LnkPrp} + T_{proc}\\
     \frac{L_C * D_{conn}}{B_w} &<& T_{LnkPrp} + T_{proc}\\
    L_C &<& \frac{(T_{LnkPrp}+ T_{proc}) \times B_w}{C_m }
    \label{equation:chunksize}
\end{eqnarray}

\begin{table}[ht]
\caption{Explanation of Notations}
\begin{center}
\begin{tabular}{|c|p{6cm}|} 
\hline  
 Notation & Explanation \\\hline  \hline
$L_H$& header size in bits\\
$L_B$& block size in bits \\
$N_C$& the number of chunks in a block \\
$L_C$& chunk size in bits \\ 
$B_w$& average bandwidth of a node\\ 
$R$  & radius of a network  \\ 
$T_P$  &  average broadcast time in a traditional blockchain network \\ 
$D_{conn}$  & average degree of connections  of a node \\ 
$T_{LinkTrans}$ & average transmission delay between the nodes
 \\ 
 $T_{LnkPrp}$ & average propagation delay between two nodes\\
$T_{ver}$ & average verification delay of a block\\
$T_{PiChu}$& average delay to propagate block in PiChu scheme\\
$T_{PH}$& average delay to propagate a header\\
$T_{DC}$& average delay in transmitting all chunks from one node to another\\ 
$T_{tc}$& transmission delay of a chunk\\
$T_{proc}$&  PiChu processing overhead delay\\\hline
\end{tabular}
\end{center}
\label{tbl:NotationExplanation}
\end{table}

\section{Defense against Potential Attacks} \label{sec:security}
In this section, we discuss the potential attacks and  mitigation strategies in a PiChu enabled blockchain network.

\subsection{Forwarding node tampers  data}
An intermediate node can modify the data in the block before forwarding it to other nodes. If a malicious node modifies the data in the chunk and forwards it, then the receiving nodes can not verify the integrity of the chunk. Receiving nodes validates the integrity of the chunk by checking the signature of the chunk. A node that received a tampered chunk discards the chunk and disconnects from the node that sent it.

The node has to receive the remaining chunks from other neighbors. A node can receive the header $H$ invitation from multiple neighbors. The node keeps a record $R$ of neighbors that sent the header $H$ invitation. When a tampered chunk is received for the block with header $H$, node disconnects from the sender and requests to pipeline remaining chunks from an optimal neighbor in the record $R$. The optimal neighbor is decided based on latency and transmission delay.

\subsection{Miner includes invalid transactions in a block}
A miner can include invalid transactions in a block. This causes one or more chunks to contain invalid transactions. 
The header will be accepted by all the nodes, as it was valid. 
When receiving the chunks, nodes validate the chunks before forwarding them.
If the chunk contains invalid transactions, then nodes can not validate that chunk. If it contains invalid transactions and integration is correct, then nodes can safely assume that the miner is malicious. After detecting that the miner is malicious, the node forwards the chunk with invalid transactions to neighbors so that other nodes can detect that miner is malicious. 

If the miner includes invalid transactions in the last chunk, then he can perform a denial of service on the network for the time it takes to broadcast the block. The time to propagate the block through PiChu is small compared to regular broadcast. So the time that the miner can perform a denial of service on the network is small. PiChu broadcasting is used on a block if adding that block to the chain makes it longer. We do this to reduce the forks in the chain and prioritize the miner that finds the block first. After detecting the invalid transactions in a chunk, nodes blacklist the header of that block. Nodes will not add a block with a blacklisted header to their blockchains. This causes the miner to lose the reward for a mined block. We can also revoke all the rewards that the miner accumulated. Miner has to lose his reward if he wants to perform a denial of service on the network.

\subsection{Intermediate node delays the sending of the chunks}
A forwarding or intermediate node in the blockchain network can intentionally delay the forwarding of chunks. The nodes connected to the attacking node receives the chunks slower than their peers. If there are many attackers in the network, then the block broadcast time will increase. The mitigation for this attack is similar to the data tampering mitigation.
The node keeps a record $R$ of neighbors that sent the header $H$ invitation. When a node detects or suspects that an intermediate is delaying the forwarding of chunks, then node disconnects from the forwarding node and requests to pipeline remaining chunks from a neighbor in the record $R$.

\subsection{Miner dies while sending the block or sends only partial block}
Miner node might fail while sending the block or intentionally sends the partial block to perform an attack on the network. It might not be possible to differentiate whether the miner node died or intentionally sent the partial block, so the approach for the two cases is the same. When a node does not receive the chunk $X$ after receiving the chunk $X-1$
, it has to first decide whether miner died or the forward node died. A forwarding node might intentionally stop forwarding the chunks. To decide either miner died or the forwarding node died , the node requests the chunk $X$ from all of its neighbors. If any of the neighbors send the chunk $X$, then the forwarding node is failed. If no neighbor sends the chunk $ X $ within a time frame, it is safe to assume that the miner died. If a forwarding node is failed, then the node terminates the connection with the forwarding node and receives the chunks from another neighbor. 

Assume that the miner died while sending the chunk $X$, and all the previous chunks are valid. All the nodes in the network will have chunks till $X-1$. Some blockchains can tolerate partial block in the chain, and other blockchains can not tolerate it. If it is tolerable, then nodes append a special chunk to the chunk $X-1$ that represents only partial block is received. After appending the special chunk, nodes will not receive any further chunks for that block. If partial blocks are not acceptable, nodes discard the chunks received for that block and keep the header in the blacklist. Nodes will not accept the block with a blacklisted header. As the header is not accepted in the blockchain, the miner will not get the reward for that block. Miner loses the reward for sending partial block. This gives the incentive to not send partial blocks. In another approach, nodes will not use the PiChu scheme for the blacklisted header and propagate through the general approach. In an aggressive approach, nodes can take away all the rewards that the miner accumulated. 

\section{Experiment Results} \label{sec:experiments}

\begin{figure*}[th]
    \centering
    \subfigure[Traditional approach]{\includegraphics[width=0.48\textwidth]{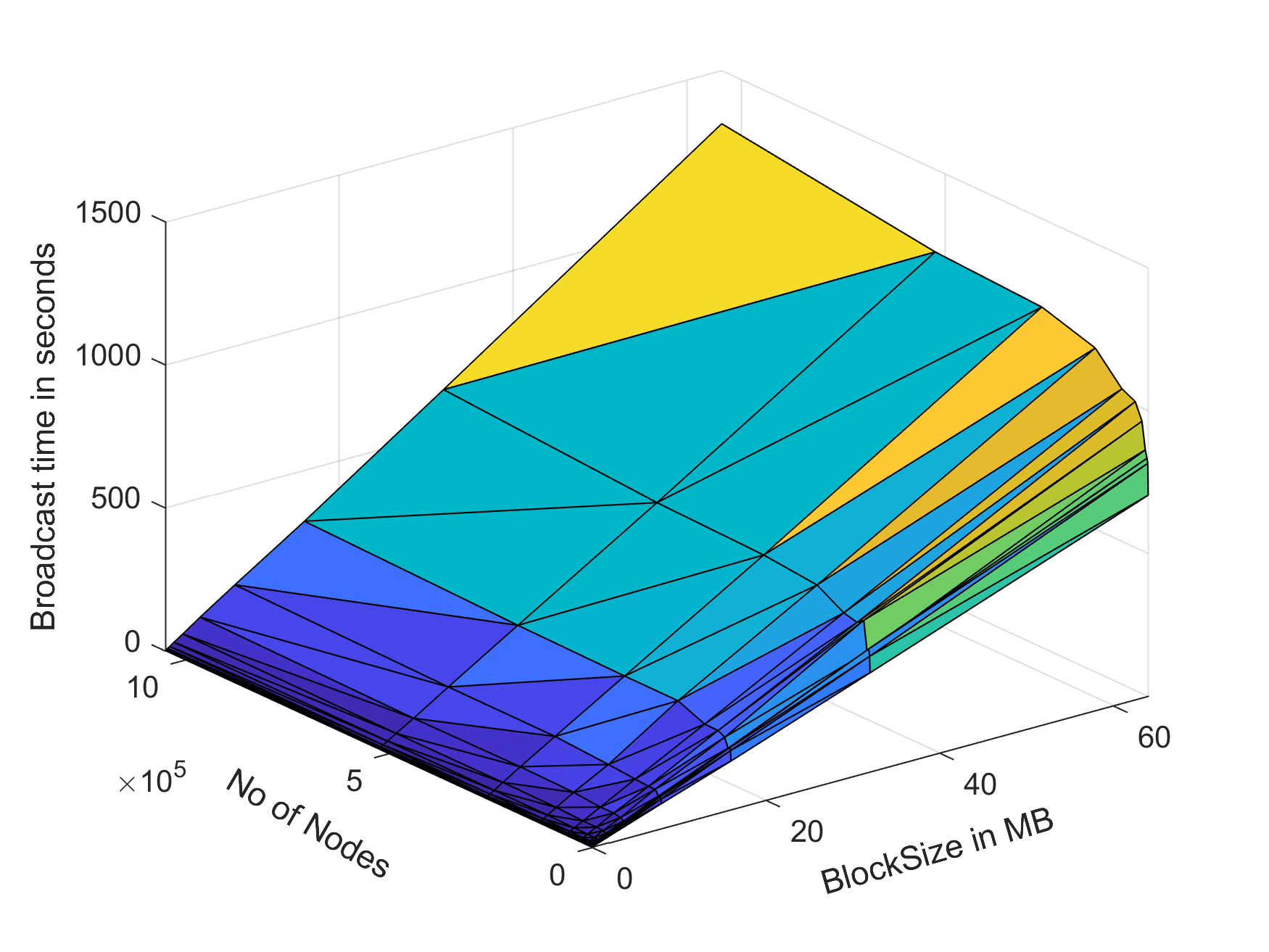}} \label{fig:regulargraph}
    \subfigure[PiChu approach]{\includegraphics[width=0.48\textwidth]{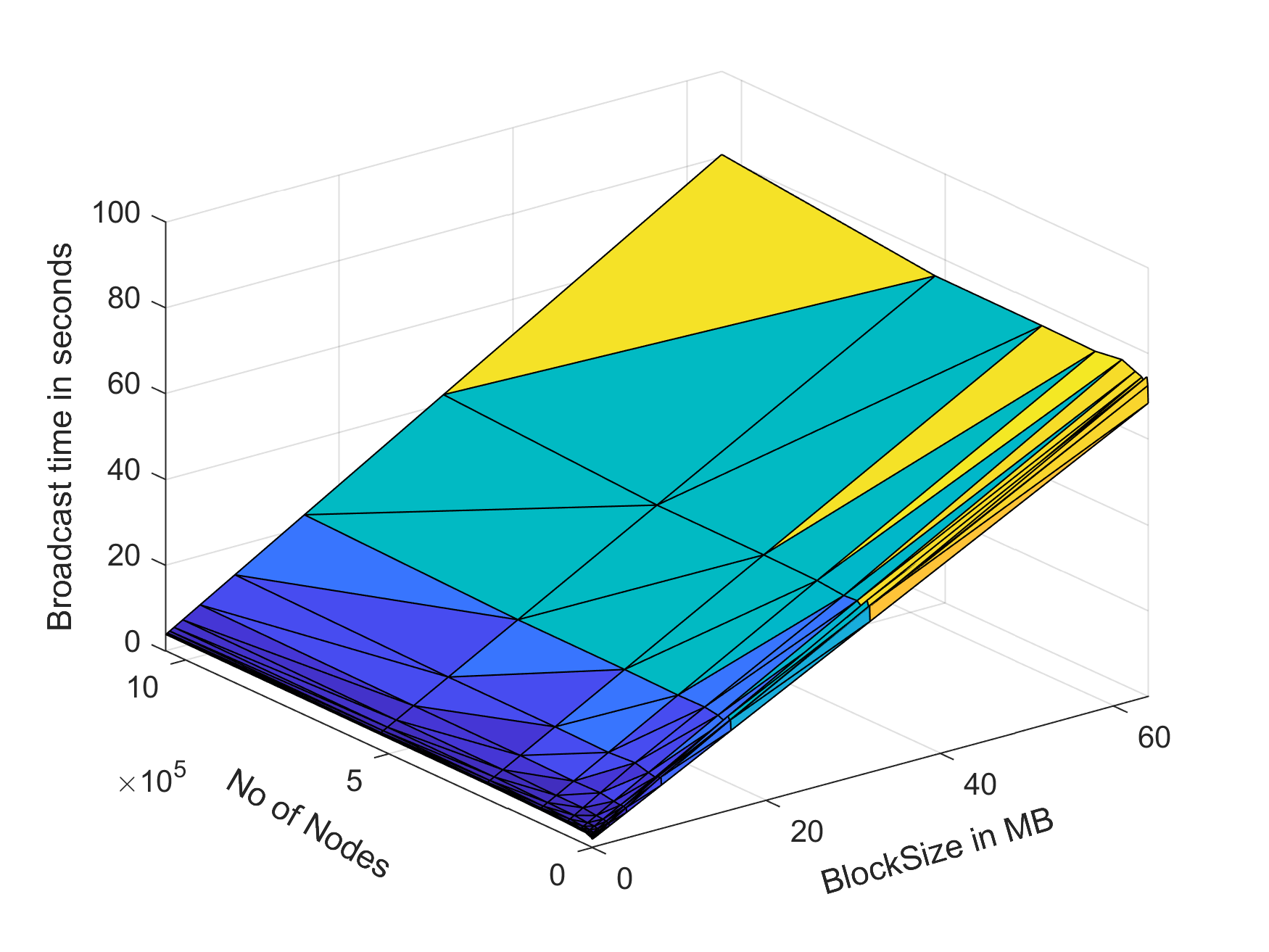}}  \label{fig:pichugraph}
    \caption{Block broadcast delay in a blockchain network  [Number of nodes: 1K $\sim$ 1M nodes; Block sizes: 8KB $\sim$  64MB]; \\(more than 15 times faster with PiChu in a million node network and 64MB blocks)}
    \label{fig:regulargraph-vs-pichugraph}
\end{figure*}

\begin{figure*}[th]
\centering
\begin{minipage}[c]{0.48\textwidth}
\begin{center}
  \includegraphics[width=\linewidth]{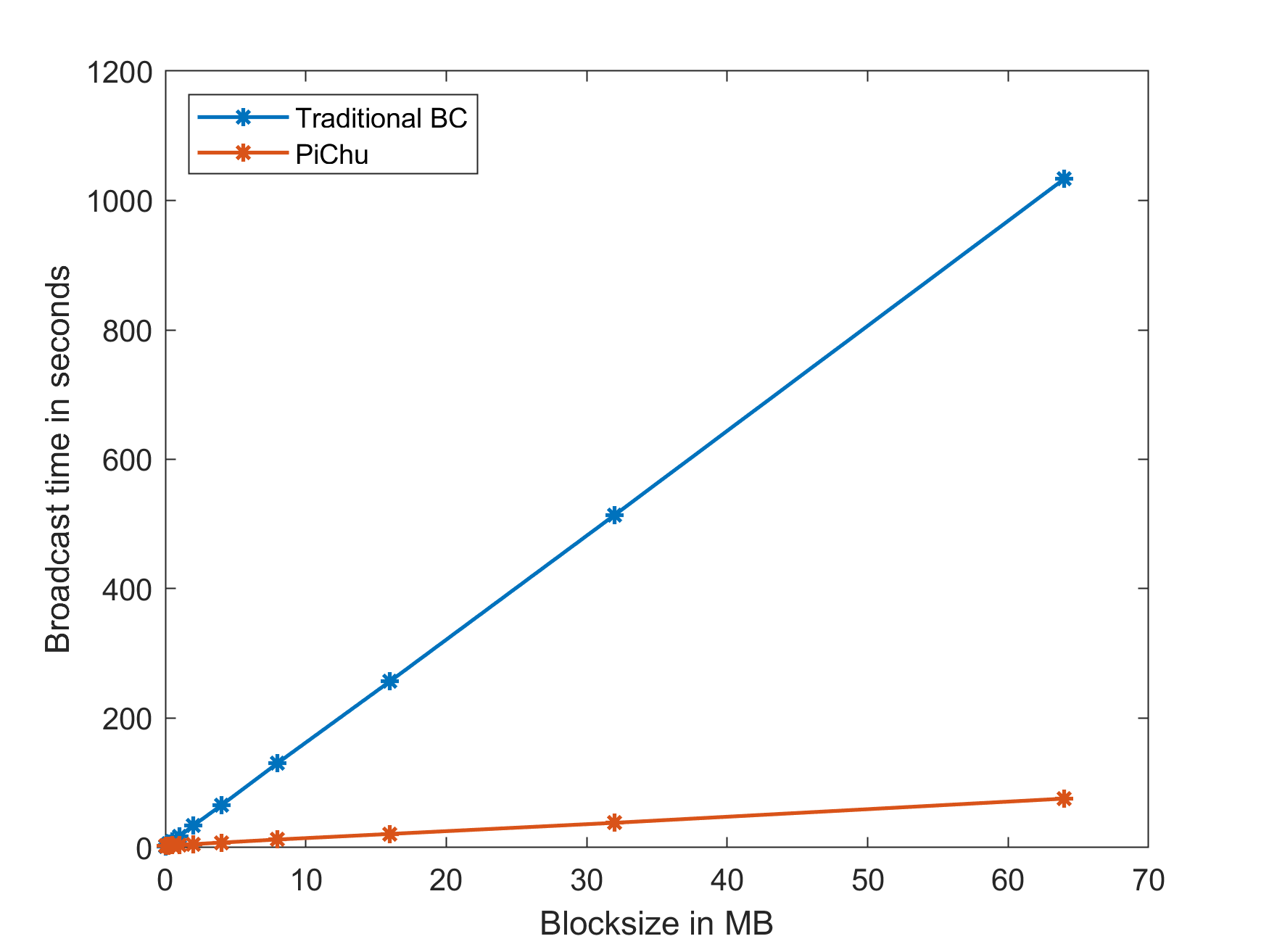}
  \caption{Block broadcast time comparison: Traditional vs. PiChu (in a 65536 node network)}
  \label{fig:comparision2d_65536}
\end{center}
\end{minipage}
\begin{minipage}[c]{0.48\textwidth}
\begin{center}
  \includegraphics[width=\linewidth]{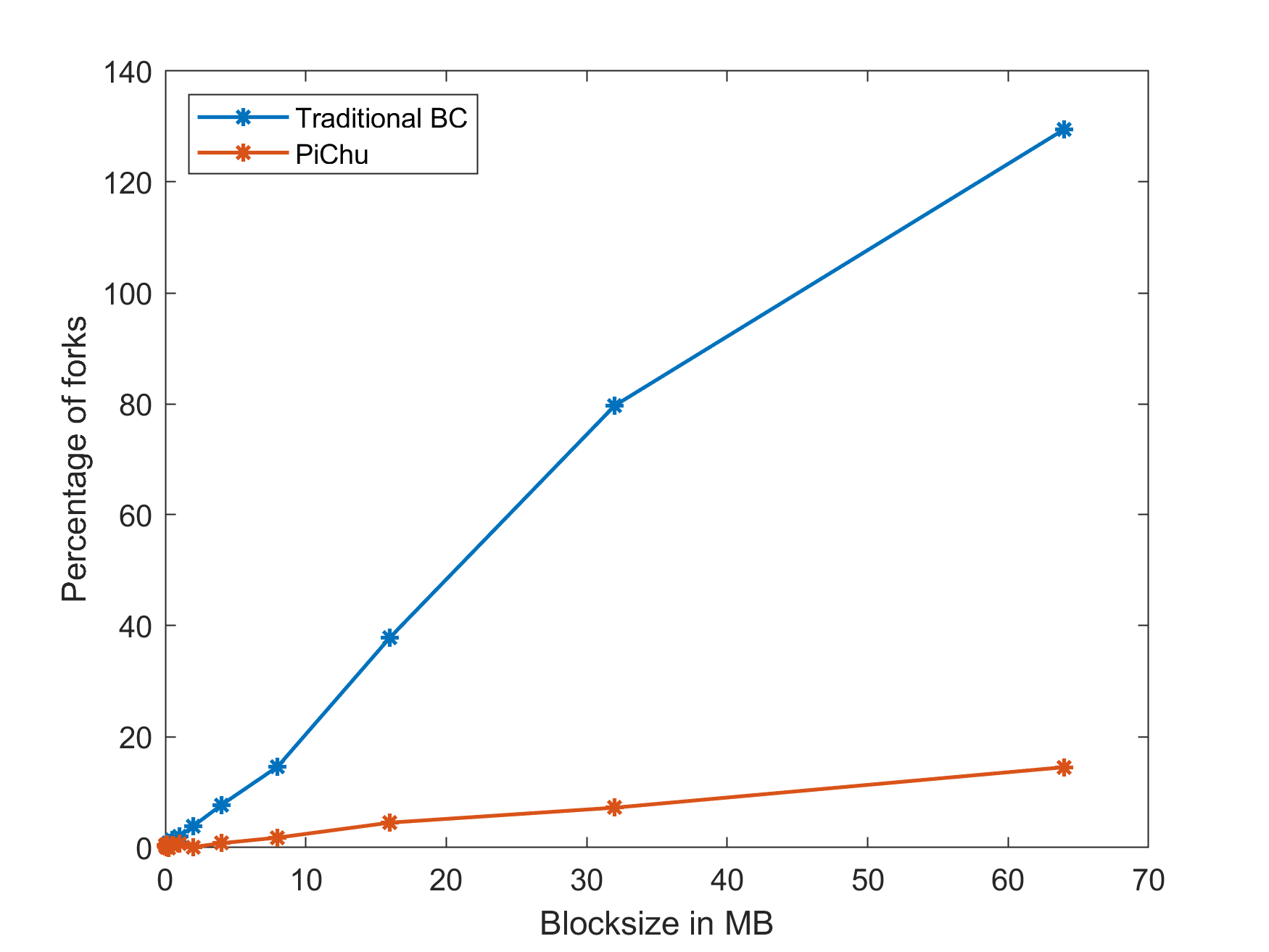}
  \caption{Percentage of forks: Traditional vs. PiChu (in a 65536	node network)}
  \label{fig:fork_comparision}
\end{center}
\end{minipage}
\end{figure*}

In order to validate the effectiveness of the PiChu scheme in a very large network with varied parameters, we have developed our own blockchain simulator. While there is an existing blockchain simulator called Simblock~\cite{simblock}, it is not well-suited to simulate the block broadcasting in a network with a large number of nodes. Our simulator is developed in Java, and we made the source code publicly available through github~\cite{BlockchainSimulator}. It can simulate block broadcasting in a network with millions of nodes and supports a large block size. Our blockchain simulator can simulate  block broadcast in traditional and PiChu approach. It takes the average bandwidth of nodes, average latency between nodes, the block size, chain length, number of nodes, and average degree of a node ($D_n$) as input. For a given number of nodes, the simulator generates a random graph topology based on the average degree per node.

We first match general propagation results with real measurements data as well as other existing simulators for comparable settings of the experiments. Table~\ref{table:simulatorsettings} shows the simulation settings used in our study that is similar to \cite{BlockPerformance} and \cite{simblock}. The output of our simulation is showed and compared in Table~\ref{table:correctness}. Our results are close to the real measurements.
\begin{table}
 	\centering
 	\caption{Simulation Settings}
 	\label{table:simulatorsettings}
 \begin{tabular}{ |c| c c c| } 
 \hline
Parameter             & Bitcoin         & Litecoin     & DodgeCoin \\
\hline \hline
\# of the nodes       & 6000            & 800          & 600       \\
Block Interval        & 10 min          & 2.5 min & 1 min     \\
Block Size            & 534 KB          & 6.11 KB      & 8 KB      \\
\# of the connections  & \multicolumn{3}{c|}{based on Miller.A\cite{coinscope}}\\
Bandwidth             & \multicolumn{3}{c|}{ testmy.net \cite{BandwidthCountries}}\\ 
Propagation delay     & \multicolumn{3}{c|}{ verizon \cite{LatencyCountries}}  \\ 
 \hline
\end{tabular}
 \end{table}

\begin{table}
\centering
\caption{Various simulators output}
\label{table:correctness}
\begin{tabular}{ |c | c | c | c| } 
 \hline
             & Bitcoin         & Litecoin     & DodgeCoin \\

(Block Interval)        & (10 m)          & (2 m 30 s) & (1 m)     \\ \hline \hline
$t_{MBP}$ of Real Measurement~\cite{BlockPerformance}      & 8.7 s  & 1.02 s & 0.98 s \\
$t_{MBP}$ from Gervais et. al.~\cite{BlockPerformance} & 9.42 s & 0.86 s & 0.83 s \\
$t_{MBP} $ from SimBlock~\cite{simblock}      & 8.94 s & 0.85 s & 0.82 s \\
$t_{MBP}$ from our Simulator     & 9.55 s     & 1.04 s    & 1.07 s  \\
\hline

Measured $r_f$   &0.41\% &0.27\%& 0.62\\
Gervais et al. $r_f$ & 1.85\% &0.24\%&0.79\%\\
SimBlock $r_f$ &0.58\%& 0.30\%& 0.80\%\\
Our Simulator $r_f$ & 0.55\% & 0.40\%&0.70\%\\
\hline
\end{tabular}

\end{table}

First,  we assess how the block broadcast delay varies with the number of nodes and the block size in the traditional broadcast scheme. 
The average bandwidth of the nodes and average latency between nodes for this experiment are taken from \cite{BandwidthCountries} and \cite{LatencyCountries}. In this experiment, the degree of each node is varied between 8 and 12. The maximum number of connections for a node in bitcoin is  125~\cite{coinscope}, and the maximum number of connections for a node in Ethereum is 50~\cite{EthereumPeers}. Coinscope\cite{coinscope} found that the majority of nodes in the Bitcoin network have a degree between 8 and 12, even though the maximum number of connections is set to 125. The verification delay is set to 0.25 ms for a transaction\cite{OnScaling}. After setting the parameters for the experiment, the number of nodes in the network and block size are varied. The results for this experiment are represented in Figure~\ref{fig:regulargraph-vs-pichugraph} (a). We can observe that when the number of nodes is constant, broadcast time increases linearly with an increase in block size. The broadcast time is proportional to the product of the network radius and block size. The broadcast time also increases with increasing the number of nodes.

We then assess how the block broadcast delay vary with the number of nodes and the block size in the blockchain networks using the PiChu propagation technique. The parameters for this experiment are the same as the previous experiment except the $D_n$. PiChu technique works better if the degree of the nodes is small. In Equation~(\ref{equation:pichuprop}), we can observe that when the block size is high, the broadcast delay majorly depends on the transmission delay of the block between two nodes. The efficiency of the PiChu depends on how fast a chunk can be transmitted from one node to another. If the degree of a node is high, then the time it takes to transmit a chunk from one node to another increases, and the efficiency of the PiChu decreases. The average degree of the node is set to 5. The reason is explained in the latter part of this section. After setting the parameters for this experiment, the number of nodes and block size are varied. Figure~\ref{fig:regulargraph-vs-pichugraph} (b) shows the output of this experiment.  We can observe that for a given number of nodes, the block broadcast time increases linearly with increasing the block size, but the slope is less than the traditional approach broadcast time. When block size is large, the propagation delay mainly depended only on block size instead of the product of the network radius and block size. The broadcast time increases a little with an increase in the network radius. 

Figure~\ref{fig:comparision2d_65536} shows how the broadcast delay change with respect to block size for 65536 nodes in PiChu and the general approach. The block propagation with PiChu for 65536 nodes is 13.6 times less than the traditional approach. The block propagation with PiChu for million nodes is 16.3 times less than the traditional approach. By the experiment results, we can say that the PiChu propagation method is efficient than traditional propagation, and the efficiency of the PiChu increases with an increase in the number of nodes or block size. 

Figure~\ref{fig:fork_comparision} shows the percentage of forks occurring with respect to block size for 65536 nodes in PiChu and traditional approach. The block interval for this experiment is 10 minutes. The percentage of forks occurring in the PiChu approach is ten times less than the traditional approach. As the forks are reducing, we can increase the size of the block. Throughput increases with an increase in block size.

\begin{figure*}[th]
\centering
\begin{minipage}[c]{0.48\textwidth}
\begin{center}
  \includegraphics[width=0.95\linewidth]{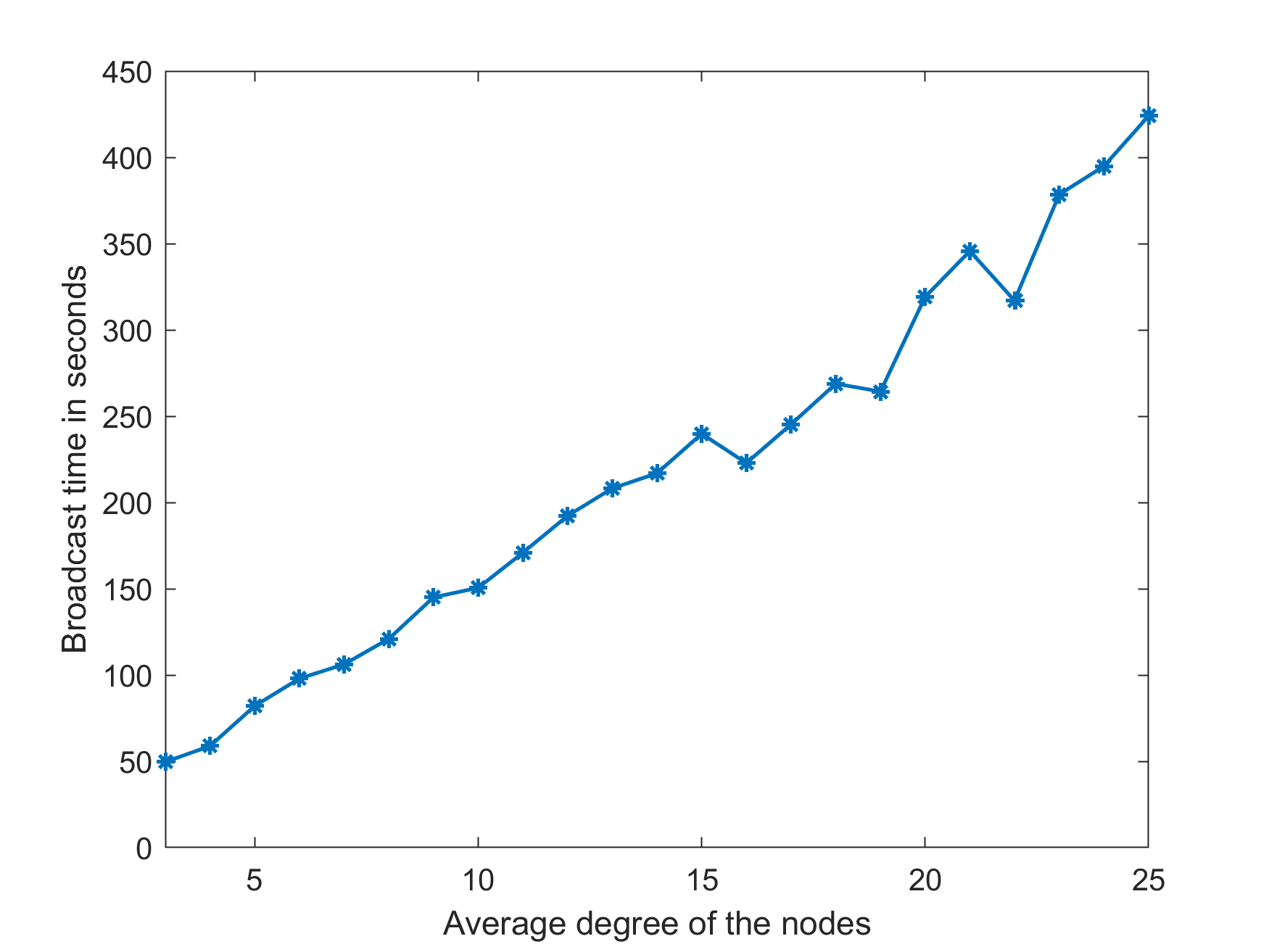}
  \caption{Block broadcast time with respect to maximum number of connections per node}
  \label{fig:cm}
\end{center}
\end{minipage}
\begin{minipage}[c]{0.48\textwidth}
\begin{center}
  \includegraphics[width=0.95\linewidth]{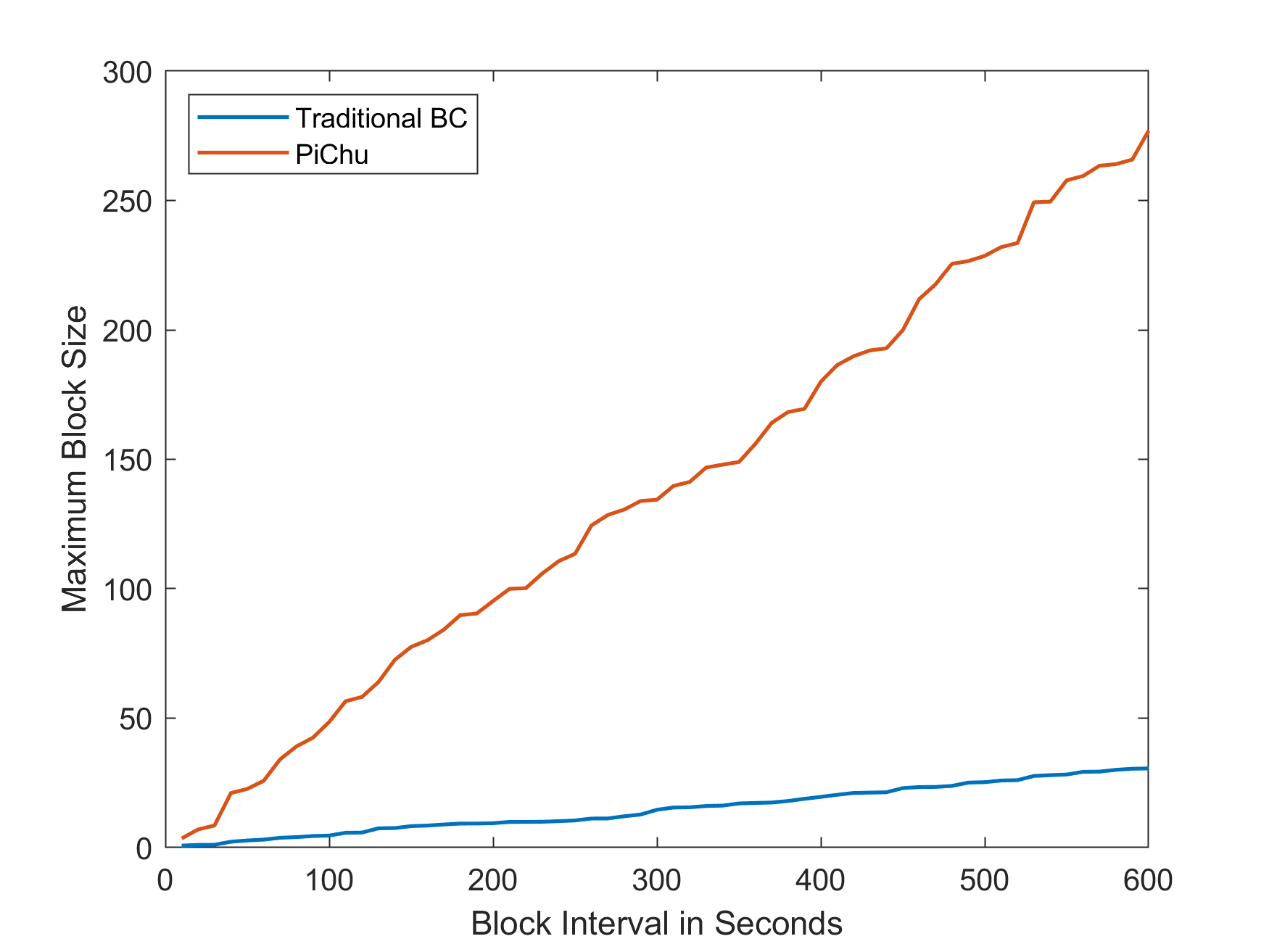}
  \caption{Maximum block size for a given interval in 65536 nodes network}
  \label{fig:maxblock}
\end{center}
\end{minipage}
\end{figure*}

The maximum possible block size for a given block interval is measured for a traditional blockchain and the PiChu blockchain. Both blockchain networks contain 65536 nodes. The maximum possible block size is for a given block interval is determined by increasing the block size until the forks are greater than or equal to 100 percent. When forks are greater than 100 percent, the blockchain becomes obsolete. Figure~\ref{fig:maxblock} shows the results. The maximum block size for a given interval is ten times higher in PiChu than traditional. In both approaches, the maximum block size increases with an increase in block interval. 

The broadcast time of a block in PiChu depends on the degree of nodes. In PiChu, the broadcast time increases with respect to the degree of  nodes as the time takes to send the chunk to all the connections increases. In the traditional approach, the broadcast time might increase or decrease with respect to node degree.  To confirm that the broadcast time increases with an increase in the node degree, we varied the node degree between 3 and 25. The degree can not be two as the topology of the network will be become linear or circular. The simulator settings are the same as the experimental settings. The number of nodes is 65536, and the block size is set to 64 MB. Figure~\ref{fig:cm} shows how the broadcast time increases with an increase in degree. The lowest broadcast time is recorded when the degree of the nodes is 3. If we choose the degree as three, then nodes are susceptible to Sybil attacks, and new nodes might find it difficult to discover the nodes. The degree should be as high as possible with reasonable broadcast time. We suggest the degree to be 5 as the block can be broadcast in under 80 seconds. It is to be noted that if we used degree 3 in our previous experiments, then the experimental results will be 1.6 times better.

\section{Conclusions} \label{sec:conclusion}

We have proposed a block acceleration scheme via pipelining and chunking of a block, named PiChu, to address the issue of scalability and performance of a blockchain network. To the best of our knowledge, this is the first kind of approach for blockchain scalability. The approach can be employed with minimal modification of existing blockchain networks. We have shown the efficiency of the PiChu approach, both theoretically and extensive evaluation using our blockchain simulator.
Our experiment results show that PiChu significantly outperforms traditional block propagation methods, and its efficiency increases with the size of a blockchain network. Our future work includes extending our blockchain simulator to support various scalability schemes and exploring the effectiveness of using  multiple scalability schemes together.

\balance
\bibliographystyle{IEEEtran}
\bibliography{IEEEabrv,./bib/cutthrough}

\end{document}